\documentclass[conference]{IEEEtran}
\usepackage{times}

\usepackage[numbers]{natbib}
\usepackage{multicol}
\usepackage[bookmarks=true]{hyperref}
\usepackage{macros}
\usepackage{paralist}
\usepackage{mathrsfs}
\usepackage[noend,linesnumbered]{algorithm2e}

\newtheorem{rema}{Theorem}
\newtheorem{remark}[rema]{Remark}
\newtheorem{theo}{Theorem}
\newtheorem{theorem}[theo]{Theorem}
\newtheorem{defi}{Theorem}
\newtheorem{definition}[defi]{Definition}

\newif\ifarxiv
\arxivtrue

\begin{document}

\title{Learning Lyapunov (Potential) Functions from Counterexamples
  and Demonstrations}


\author{\authorblockN{Hadi Ravanbakhsh and Sriram Sankaranarayanan}
\authorblockA{Department of Computer Science, University of Colorado Boulder, Boulder, Colorado 80302\\
Email: firstname.lastname@colorado.edu}
}


%

\maketitle

\begin{abstract}
  We present a technique for learning control Lyapunov (potential)
  functions, which are used in turn to  synthesize controllers
  for nonlinear dynamical systems.  The learning framework uses a
  \emph{demonstrator} that implements a black-box, untrusted strategy
  presumed to solve the problem of interest, a \emph{learner} that
  poses finitely many queries to the demonstrator to infer a candidate
  function and a \emph{verifier} that checks whether the current
  candidate is a valid control Lyapunov function. The overall learning
  framework is iterative, eliminating a set of candidates on each
  iteration using the counterexamples discovered by the verifier and
  the demonstrations over these counterexamples. We prove its
  convergence using ellipsoidal approximation techniques from convex
  optimization. We also implement this scheme using nonlinear MPC
  controllers to serve as demonstrators for a set of state and
  trajectory stabilization problems for nonlinear dynamical systems. Our
  approach is able to synthesize relatively simple polynomial control
  Lyapunov functions, and in that process replace the MPC using a
  guaranteed and computationally less expensive controller.

\end{abstract}

\IEEEpeerreviewmaketitle

\section{Introduction}\label{sec:introduction}

We propose a novel \emph{learning from demonstration} scheme for
inferring control Lyapunov functions (potential functions) for
stabilizing nonlinear dynamical systems to reference states/
trajectories.  Control Lyapunov functions (CLFs) have wide
applications to motion planning problems in robotics. They extend the
classic notion of Lyapunov functions to systems involving control
inputs~\cite{artstein1983stabilization}. Finding a CLF also leads us
to an associated feedback control law that can be used to solve the
stabilization problem. Additionally, they can be extended for feedback
motion planning using extensions to time-varying or sequential
CLFs~\cite{burridge1999sequential,tedrake2010lqr}. Likewise, they have
been investigated in the robotics community in many forms including
\emph{artificial potential functions} to solve path planning problems
involving obstacles~\cite{lopez1995autonomous}.

However, synthesizing CLFs for nonlinear systems remains a
challenge~\cite{primbs1999nonlinear,rantzer2001dual}. Standard
approaches to finding CLFs include the use of dynamic programming,
wherein the value function satisfies the conditions of a 
CLF~\cite{bertsekas1995dynamic}, or using non-convex bilinear matrix
inequalities (BMI)~\cite{henrion2005solving}. BMIs can be solved using
alternating minimization
methods~\cite{el1994synthesis,tan2004searching,majumdar2013control}. However,
these approaches often get stuck in local minima and exhibit poor
convergence guarantees~\cite{Helton+Merino/1997/Coordinate}.

In this article, we investigate the problem of learning a CLF using a
black-box \emph{demonstrator} that can be queried with a given system
state, and responds by demonstrating control inputs to stabilize the
system starting from that state.  However, our framework uses just the
control input at the query state.  Such a demonstrator can be realized using an
expensive nonlinear model predictive controller (MPC) that uses a local optimization scheme,
or even a human operator under certain assumptions~\footnote{We do not
  handle noisy or erroneous demonstrations in this
  paper}.  The framework has a \textsc{Learner} which selects a
candidate CLF and a \textsc{Verifier} that tests whether this CLF is
valid. If the CLF is invalid, the \textsc{Verifier} returns a state at
which the current candidate fails. The \textsc{Learner} queries the
demonstrator to obtain a control input corresponding to this state.
It subsequently eliminates the current candidate along with a set of related
functions from further consideration. The framework continues to
exhaust the space of candidate CLFs until no CLFs remain or a valid
CLF is found in this process.

We prove the process can converge in finitely many steps provided the
\textsc{Learner} chooses the candidate function appropriately at each
step. We also provide efficient SDP-based approximations to the
verification problem that can be used to drive the framework. Finally,
we test this approach on a variety of examples, by solving
stabilization problems for nonlinear dynamical systems. We show that
our approach can successfully find simple CLFs using finite horizon
nonlinear MPC schemes with appropriately chosen cost functions to
serve as demonstrators. In these instances, the CLFs yield control
laws that are computationally inexpensive, and  guaranteed against
the original dynamical model.



\subsection{Illustrative Example: \textsc{Tora} System}
\begin{figure*}[t]
\begin{center}
\includegraphics[width=0.95\textwidth]{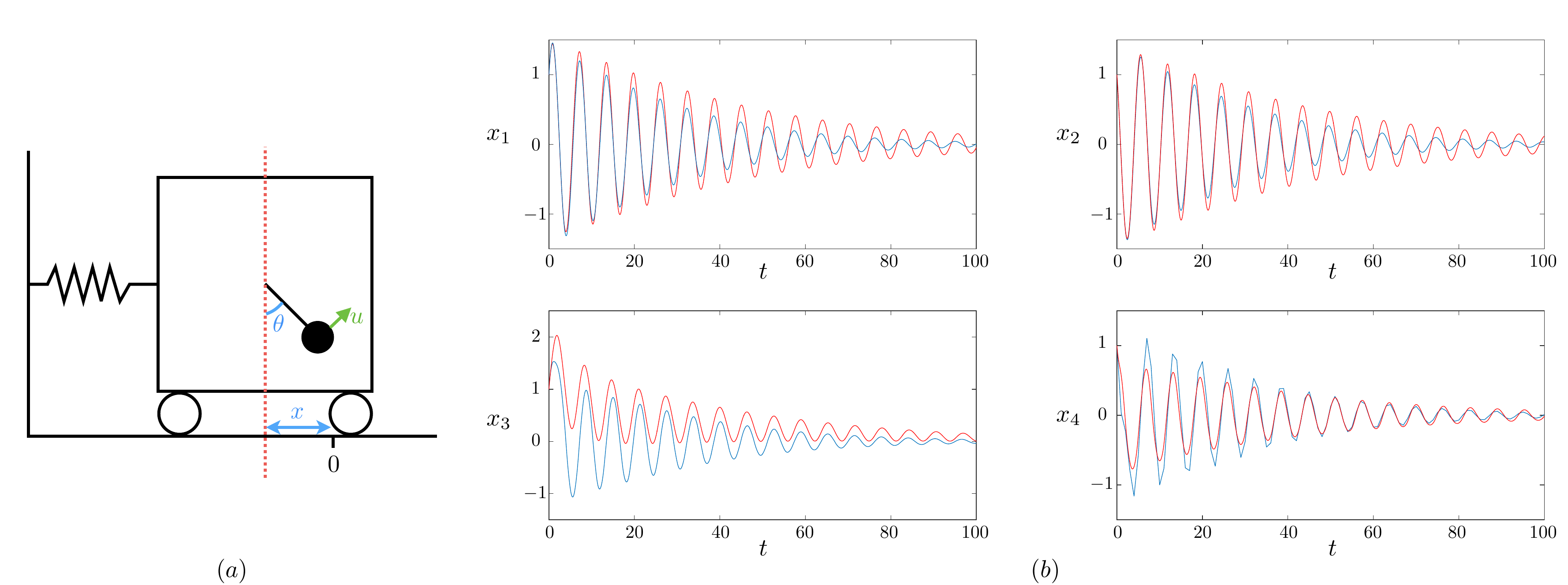}
\end{center}
\vspace*{-0.4cm}
\caption{Tora System. (a) A schematic diagram of the \textsc{Tora} 
system. (b) Execution traces of the system using MPC control 
(blue traces) and Lyapunov based control (red traces) starting 
from same initial point.}
\label{Fig:tora-example}
\vspace*{-0.4cm}
\end{figure*}

Figure~\ref{Fig:tora-example}(a) shows a mechanical system consisting of
a cart attached to a wall using a spring. The position of the cart $x$
is controlled by an arm with a weight that can be moved back and forth
by applying a force $u$, as shown. The goal is to stabilize the cart
to $x=0$, with its velocity, angle, and angular velocity
$\dot{x} = \theta=\dot{\theta} = 0$. We refer the reader to the Jankovic et
al.~\cite{jankovic1996tora} for a derivation of the dynamics shown below
in terms of state variables $(x_1,\ldots,x_4)$ and 
control input $u_1$, after a suitable change of basis transformation:
\begin{equation}\label{eq:tora-dyn}
		\dot{x_1} = x_2,\, \dot{x_2} = -x_1 + \epsilon \sin(x_3),\, \dot{x_3} = x_4,\, \dot{x_4} = u_1\,.
\end{equation}
We approximate $\sin(x_3)$ using a degree $3$ approximation which is
quite accurate over the range $x_3 \in [-2,2]$.
The equilibrium $x = \dot{x} = \theta = \dot{\theta} = 0$ now
corresponds to $x_1 = x_2 = x_3 = x_4 = 0$.  The state space is taken
to be $S: [-1,1] \times [-1,1] \times [-2,2] \times [-1,1]$, the control input 
$u_1 \in [-1.5, 1.5]$. 

\paragraph{MPC Scheme:} A first approach to doing so uses a nonlinear
model-predictive control (MPC) scheme using
the time horizon $\T = 30$, time step $\tau = 1$
and a simple cost function
\[
  \sum_{t=[0,\tau,...,\T]} \left(||\vx(t)||_2^2 +
    ||\vu(t)||_2^2\right) + \T \ ||\vx(\T)||_2^2 \,.\] Given the
model in~\eqref{eq:tora-dyn}, such a control is implemented using a
first order numerical gradient descent method to minimize the cost
function over a finite time horizon.  The convergence was informally
confirmed by observing hundreds of such simulations from different
initial conditions for the system. However, the MPC scheme is
expensive, requiring repeated solutions to (constrained) nonlinear
optimization problems in real-time. Furthermore, the closed loop lacks
formal guarantees despite the \emph{high confidence} gained from
numerous simulations.

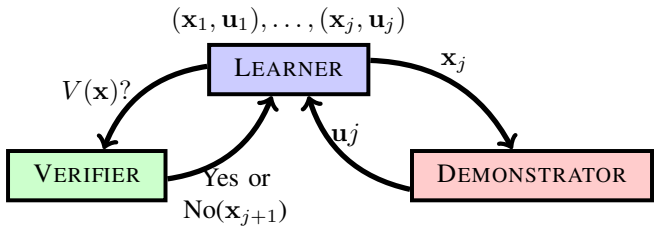
\begin{figure}[t]
\begin{center}
\begin{tikzpicture}
\matrix[every node/.style={rectangle, draw=black, line width=1.5pt}, row sep=20pt, column sep=15pt]{
  & \node[fill=blue!20](n0){\begin{tabular}{c} 
\textsc{Learner} \end{tabular}}; & \\
\node[fill=green!20](n1){\begin{tabular}{c}
\textsc{Verifier} \end{tabular}}; & & \node[fill=red!20](n2){\begin{tabular}{c}
\textsc{Demonstrator} \end{tabular} }; \\
};

\path[->, line width=2pt] (n0) edge[bend right] node[left]{$V(\vx)?$} (n1)
(n1) edge[bend right] node[below]{\;\begin{tabular}{c}
Yes or \\
No($\vx_{j+1}$)\end{tabular}} (n0)
(n0) edge [bend left] node[above]{$\vx_j$} (n2)
(n2) edge [bend left] node[above]{$\vuـj$} (n0);
\draw (n0.north)+(0,0.3cm) node {$(\vx_1, \vu_1), \ldots, (\vx_j, \vu_j)$};
\end{tikzpicture}
\end{center}
\vspace*{-0.3cm}
\caption{Overview of the learning framework for learning a Lyapunov function.}\label{fig:learning-framework}
\vspace*{-0.3cm}
\end{figure}

\paragraph{Learning a Control Lyapunov Function:} The approach in this paper
uses the MPC scheme as a ``\textsc{demonstrator}'', and attempts to learn a
simpler control law through a control Lyapunov function. The key idea
depicted in Fig.~\ref{fig:learning-framework} is to pose queries to
the MPC at finitely many \emph{witness} states
$W = \{ \vx_1, \ldots, \vx_j \}$ and use the corresponding
instantaneous control inputs $\vu_1, \ldots, \vu_j$, respectively.  The
\textsc{learner} attempts to find a candidate function
$V(\vx)$ that is positive definite and which decreases at each witness
state $\vx_j$ through the control input $\vu_j$.  
This function is fed to the \textsc{verifier}, which checks
whether $V(\vx)$ is indeed a CLF, or discover a state $\vx_{j+1}$ at
which the condition fails. This new state is added to the witness set
and the process is iterated.

The procedure described
in this paper synthesizes the control Lyapunov 
function after $60$ iterations
of the learning loop and synthesizes the CLF $V(\vx)$ below:
\[ {\small \left(\begin{array}{c} 1.22  \ x_2^2  + 0.31  \ x_2x_3 \  + 0.44  \ x_3^2  - 0.28  \ x_4x_2 + 0.8  \ x_4x_3 \\
+ 1.69  \ x_4^2  + 0.069  \ x_1x_2  - 0.68  \ x_1x_3 - 1.85  \ x_4 x_1  + 1.60  \ x_1^2\end{array}\right)} \]

This function yields a simple associated control law that can be
implemented and guarantees the stabilization of the
model~\eqref{eq:tora-dyn}.
Figure~\ref{Fig:tora-example}(b) shows a closed loop trajectory of
this control law vs control law extracted by MPC. The advantage of
this law is that its calculation is \emph{much simpler}, and furthermore, the control is formally
guaranteed, at least for the model of the system.

\paragraph{Contributions:} In this paper, we instantiate the learning
scheme sketched above, and show that under suitable assumptions
terminates in finitely many iterations to either yield a control Lyapunov
function $V(\vx)$ that is guaranteed to be valid, or show that Lyapunov
function of a specific form does not exist.
We demonstrate this scheme and its scalability on several interesting
vehicle dynamics taken from the literature to solve stabilization to state and
trajectory stabilization problems.

\section{Background}\label{sec:background}

In this section, we briefly describe control Lyapunov 
(potential) functions. A state feedback control system
$\Psi(X, U, \Plant, \Ctrl)$ consists of a plant $\Plant$ and a
controller $\Ctrl$ over state space $X \subseteq \reals^n$ and input
space $U \subseteq \reals^m$.  The plant $\Plant$ has a state
$\vx \in X$ and input $\vu \in U$. The vector field for the plant is
defined by a smooth function $f: X \times U \rightarrow
\reals^n$. Throughout the paper, we consider control affine systems
that are possibly nonlinear in $\vx$, but affine in $\vu$, of the form

\begin{equation} \label{eq:control-affine}
	\dot{\vx} = f(\vx, \vu) = f_0(\vx) + \sum_{i=1}^m f_i(\vx) u_i \,,
\end{equation}
where $f_i: X \rightarrow \reals^n$.
The controller $\Ctrl$ measures the state of the plant ($\vx \in X$)
and provides feedback $\vu \in U$. The controller is defined by a
continuous feedback function $\K:X \rightarrow U$.
\begin{center}
\includegraphics[width=0.2\textwidth]{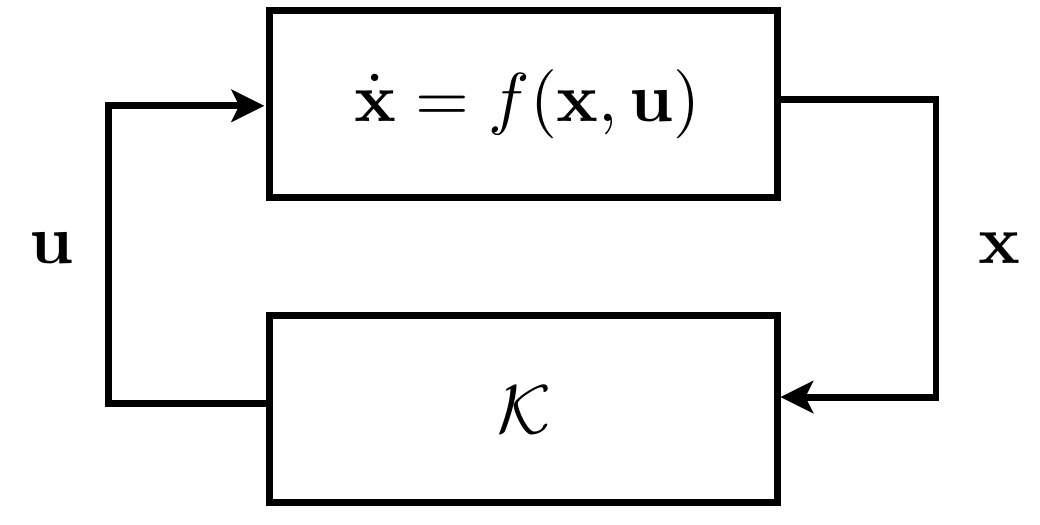}
\end{center}
For a given feedback function $\K$, the execution trace of the system
$\Psi$ is defined as $\vx(.) : \reals^+ \rightarrow X$, which maps
time to state. Formally, given $\vx(0) = \vx_0$, $\vx(.)$ is defined
according to $\dot{\vx}(t) = f_0(\vx(t)) + \sum_{i=1}^m f_i(\vx(t))(\K(\vx(t)))_i$,
where $\dot{\vx}(.)$ is the right derivative of $\vx(.)$.

In this article, we study stabilization of nonlinear systems using
Lyapunov functions (or potential function) inside a compact set
$S$. More precisely, we consider a compact and connected set
$S \subset X$. Without loss of generality, the origin $\vzero$ is the
state we seek to stabilize to. Furthermore, $\vzero \in int(S)$. We
restrict the set $S$ to be a basic semi-algebraic set defined by a
conjunction of polynomial inequalities. Likewise, the control inputs
$U$ are restricted to a polytope. 



\begin{figure}[!t]
\begin{center}
	\includegraphics[width=0.3\textwidth]{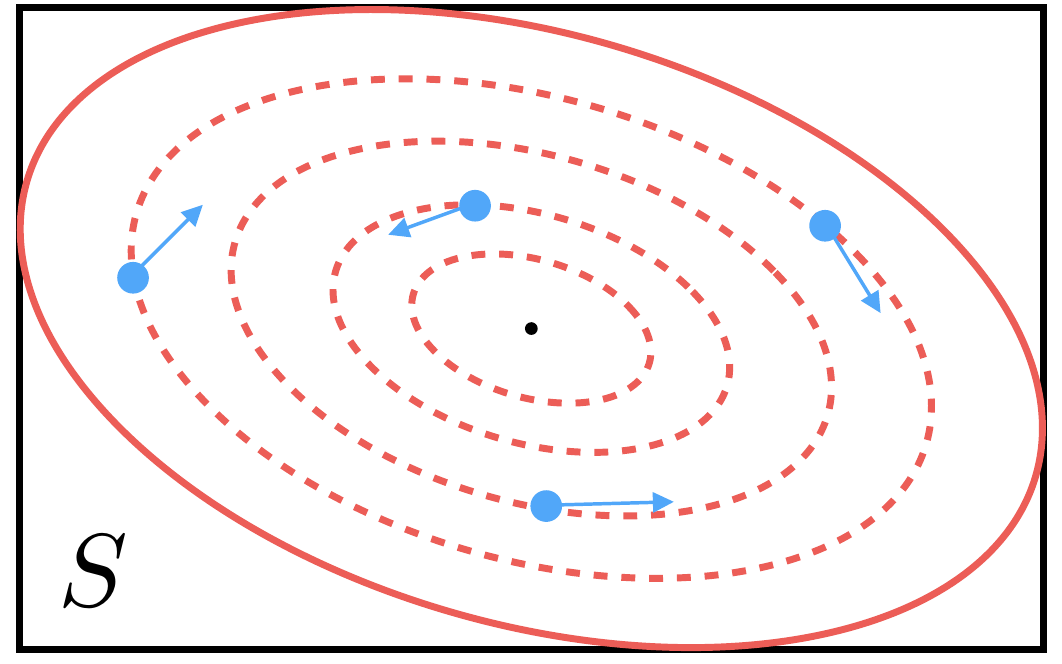}
\end{center}
\caption{Local Control Lyapunov Function (CLF): Level-sets of a CLF
  $V$ are shown using the red lines. For each state (blue dot), the
  vector field $f(\vx, \vu)$ for $\vu = \K(\vx)$ is the blue arrow,
  and it points to a direction which decreases $V$. The $\beta^*$-level
  set of $V$ ($V^{=\beta^*}$) is shown as a solid red line.}\label{fig:clf}
\end{figure}

Our approach relies on control Lyapunov functions. A \emph{control
  Lyapunov function} (CLF) ~\cite{artstein1983stabilization} $V$,
is a continuous function that respects the following conditions:
\begin{equation}\label{eq:clf-def} 
\begin{array}{rl}
& V(\vzero) = 0 \\
  (\forall \vx \in S \setminus \{0\}) \ & V(\vx) > 0 \\
	(\forall \vx \in S \setminus \{0\}) \ (\exists \vu \in U)\ & \nabla V \cdot f(\vx, \vu) < 0 \,. \\ 
\end{array}
\end{equation}
The last condition ensures that the value of $V$ can be decreased
everywhere by choosing a proper feedback $u$. Let
\[ V^{\Join \beta} = \{\vx | V(\vx) \Join \beta \}, \mbox{where}\ \Join \in \{ =, \leq, <, \geq, > \} \,. \]
Let $\beta^*$ be maximum $\beta$ s.t. $V^{\leq \beta} \subseteq S$.
Once a CLF is obtained, it guarantees that the system initialized to 
any state belonging to $V^{< \beta^*}$, can be
stabilized to the origin (Fig.~\ref{fig:clf}).  A control Lyapunov
function guarantees that there is a control strategy, which stabilizes
the system. 

\begin{theorem}
	Given a control affine system $\Psi$, where $U : \reals^m$ 
	and a polynomial control Lyapunov function $V$ satisfying Eq.~\eqref{eq:clf-def},
	there is a feedback function $\K$ for which
	if $\vx_0 \in V^{< \beta^*}$, then:
	\begin{enumerate}
		\item $(\forall t \geq 0) \ \vx(t) \in S$
		\item $(\forall \epsilon > 0) \ (\exists T \geq 0) \ \norm{\vx(T) - \vzero} < \epsilon$\,.
	\end{enumerate}
\end{theorem}

\ifarxiv
\begin{proof}
	1) Using results from Artstein~\cite{artstein1983stabilization}, 
	there exists a feedback function $\K^*$ s.t. while $\vx \in S$, then 
	$\frac{dV}{dt} = \nabla V \cdot f(\vx, \vu) < 0$. Assuming $\vx(0) = \vx_0 \in V^{<\beta^*} \subset S$, then initially $V(\vx(0)) < \beta^*$. Now, assume the state reaches $\partial S$ at time $t_2$. By continuity of $V$, there is a time $0 < t_1 \leq t_2$ s.t. $V(\vx(t_1)) = \beta^*$ and $(\forall t \in [0, t_1]) \ \vx(t) \in S$. Therefore
	\[
	V(\vx(t_1)) = \left(V(\vx(0)) + \int_{0}^{t_1} \frac{dV}{dt} dt\right) < V(\vx(0)) \,.
	\]
	This means $V(\vx(t_1)) < \beta^*$, which is a contradiction. Therefore, the state never reaches $\partial S$ and remains in $int(S)$ forever.
	2) $V$ would be a Lyapunov function for
	the closed loop system when the control unit is replaced with the feedback function $\K^*$ and using standard results in Lyapunov theory
	$(\forall \epsilon > 0) \ (\exists T \geq 0) \ ||\vx(T) - 0|| < \epsilon$.
\end{proof}
\fi

Given, a control Lyapunov function, it is possible to then obtain a
feedback law in a closed form that stabilizes the system.
Sontag~\cite{sontag1989universal} provides a method for extracting a
continuous feedback function $\K$ for control affine systems from a
control Lyapunov function. This can be extended to systems with
constraints on the control inputs~\cite{malisoff2000universal}. 
Also, feedback synthesis for periodic time/event
triggered switching is possible~\cite{ravanbakhsh2015counter,curtis2003clf}.

We have thus far considered the problem of stabilizing to a fixed
equilibrium state. However, given this primitive, we can extend the
CLF approach to related problems of (a) \emph{Reach-while-stay:}
reaching a given target set of states $T$ starting from an initial set
$I$, while staying inside a safe set 
$S$ using Lyapunov-barrier functions~\cite{prajna2004safety,ravanbakhsh2016robust};
 (b) \emph{Trajectory stabilization:} stabilizing to a trajectory 
 $\vx(t)$ rather than to a fixed state using time-varying Lyapunov 
 functions; or similarly, (c) \emph{Feedback motion planning} which
 addresses the robustness of a plan using 
 funnels~\cite{burridge1999sequential,tedrake2010lqr} ; 
 (d) \emph{Obstacles:} problems that involve reaching
while avoiding an obstacle region in the 
state-space using artificial potential functions~\cite{lopez1995autonomous}.
We will focus our exposition on the basic formulation for stability (Eq.~\eqref{eq:clf-def})
while demonstrating extensions to some of other applications mentioned above
through numerical examples.
 
%


\section{Algorithmic Learning Framework}

Finding CLFs is known to be a hard problem, requiring the solution
to BMIs~\cite{tan2004searching} or hard polynomial 
constraints\cite{ravanbakhsh2015counter}. 
A standard approach to discovering
such functions is to choose a set of basis functions
$g_1, \ldots, g_r$ and search of a function of the form
\begin{equation} \label{eq:clf-template}
	V_\vc(\vx) = \sum_{j=1}^r c_j g_j(\vx) \,,
\end{equation} 
where $\vc \in \reals^r$ is vector of unknowns. One possible choice of
basis functions involves monomials $g_j(\vx):\ \vx^{\alpha_j}$
wherein $|\alpha_j|_1 \leq D_L$ for some degree bound $D_L$ for the learning
concept (CLF). Then, the
problem is reduced to finding $\vc$ s.t. $V_\vc$ satisfies Eq.~\eqref{eq:clf-def}.

We now present the algorithmic learning framework.
Let us fix a control affine system $\Plant$ over a state-space $X$,
control inputs $U$ given by~\eqref{eq:control-affine}.
Let $\vx^* = \vzero$ be the
equilibrium we wish to stabilize the system to,
while remaining inside $S \subset X$. 

Next, we assume a \textsc{demonstrator} as a function $\D: S \mapsto U$
that given a state $\vx \in S$, provides us an appropriate feedback
$\D(\vx) \in U$ for the state $\vx$, such that $\D$ is presumed to be
a valid function that stabilizes the system. 

\begin{remark}  
  Our definition of a demonstrator is general enough to allow offline
  MPC, sample based
  methods~\cite{lavalle2000rapidly,kocsis2006bandit}, human operator
  demonstrations~\cite{khansari2017learning}, or even demonstrations that rely on opaque models
  such as neural networks.


  Also, we assume that the demonstrator is \emph{presumed}
  correct. However, the approach can work even if the demonstrator may
  fail on some input states. Finally, a faulty demonstrator may, at
  the worst, lead our technique to fail without finding a CLF. In
  particular, such a demonstrator will not cause our technique to synthesize an
  incorrect CLF.
\end{remark}

\begin{definition}[Problem Statement]\label{def:problem-stmt}
  The CLF learning problem has the following inputs:
\begin{compactenum}
  \item  A dynamical system $\Plant$ in the form
  ~\eqref{eq:control-affine}, 
 \item A safe set $S$, 
\item A ``black-box'' demonstrator function $\D: S \mapsto U$ that
  presumably stabilizes the system, and 
\item A candidate space for CLFs
  of the form $\sum_{j=1}^r c_j g_j(\vx)$ given by basis functions
  $\vg(\vx):\ \tupleof{g_1(\vx),\ldots, g_r(\vx)}$ and a compact set
  $C \ni (c_1, \ldots, c_r)$. We represent the coefficients
  $(c_1, \ldots, c_r)$ collectively as $\vc$.
\end{compactenum}

The output can be \textsc{Success:} a function $V_c(\vx): \vc^t \cdot \vg(\vx)$ that is a CLF; or
\textsc{Failure:} no function could be discovered by our procedure.
\end{definition}

\subsection{Algorithmic Learning Framework}

\begin{figure}[t]
\begin{center}
\includegraphics[width=0.40\textwidth]{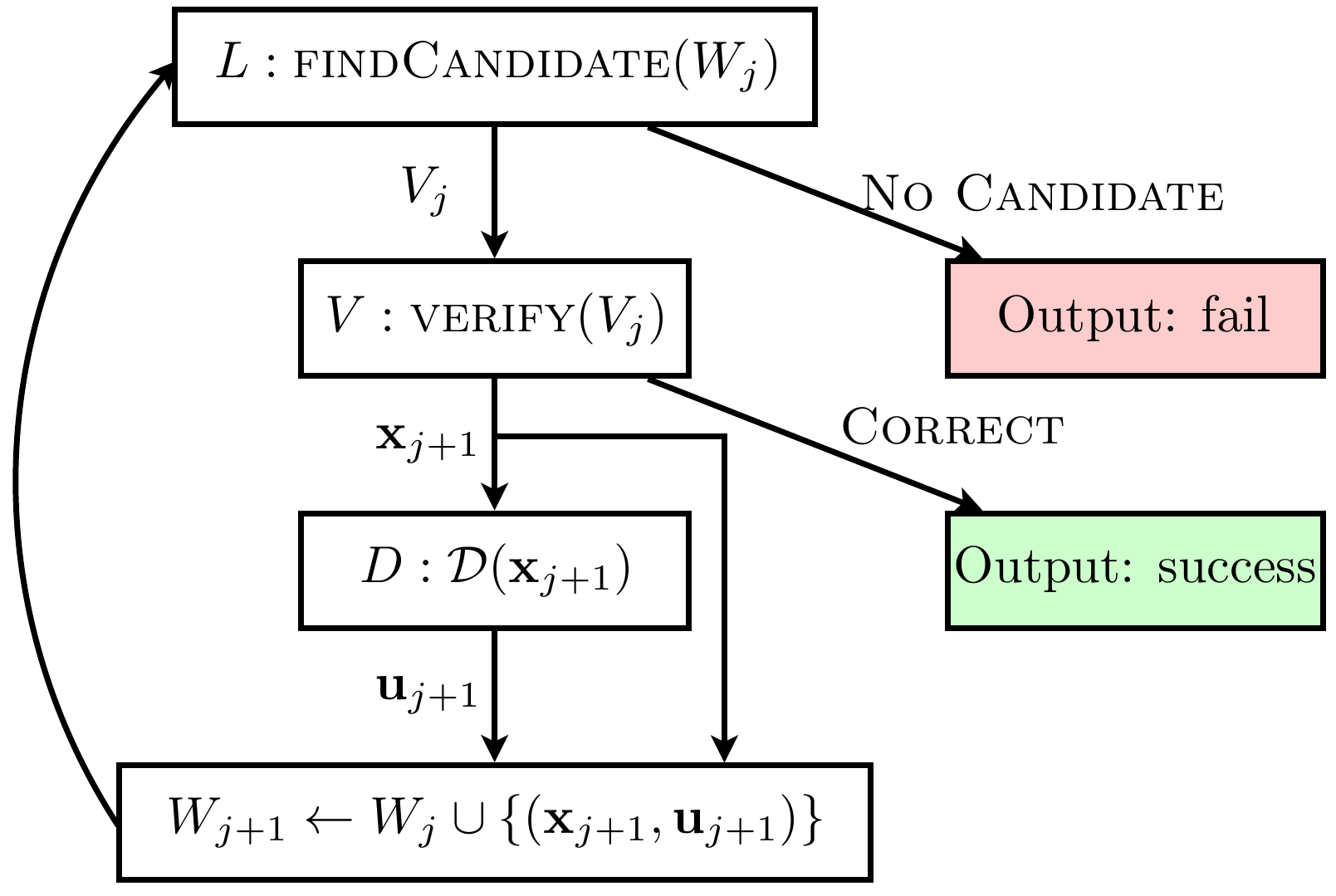}
%
%
\end{center}
\caption{Flowchart for the algorithmic learning framework. L: Learner, V: Verifier and D: Demonstrator. } \label{fig:algo-learn-flow}
\end{figure}

The algorithmic learning framework is shown in
Fig.~\ref{fig:algo-learn-flow}, and implements two modules (a)
\textsc{Learner} and (b) \textsc{Verifier} that interact with each
other and the demonstrator.  The framework works iteratively until
termination. At the $j^{th}$ iteration, the learner maintains a (witness) set
\[ W_j: \{ (\vx_1, \vu_1) ,\ldots, (\vx_{j}, \vu_j) \} \subseteq S \times U \,. \] 
$W_j$ is a finite set of pairs of states $\vx_i$ and corresponding 
demonstrated feedback $\vu_i$.
Corresponding to $W_j$, $C_j \subseteq C$ is defined as set of candidate 
coefficients  for functions $V_{\vc}(\vx): \vc^t \vg(\vx)$ with  $\vc  \in C_j$. 
Formally, $C_j$ is a set of all candidates $\vc$ s.t. $V_\vc$ satisfies the CLF 
condition~\eqref{eq:clf-def} for every point in the finite set $W_{j}$:
\begin{equation} \label{eq:C_j}
C_j : \left\{ \vc \in C\ \left|\ \mathop{\bigwedge}_{(\vx_i, \vu_i) \in W_j}
\begin{array}{c}\ V_\vc(\vx_i) > 0\ \land\ \\ \nabla V_c \cdot f(\vx_i, \vu_i) < 0 \end{array}\ \right.\right\} \,.
\end{equation}

The flowchart for the overall procedure is shown in Fig.~\ref{fig:algo-learn-flow}.

To begin with, $W_0 : \emptyset$ and $C_0: C$.  Each iteration works
as follows:
\begin{compactenum}
\item The learner samples a value $\vc_j \in C_j$ and outputs the corresponding
function $V_j(\vx): \vc_j^t \cdot \vg(\vx)$. If $C_j = \emptyset$ then no sample is found and the algorithm fails. 
\item The verifier checks if $V_j$ is a CLF by checking the conditions in ~\eqref{eq:clf-def}. If $V_j$ satisfies the conditions, then the algorithm stops to declare success. Otherwise the verifier selects a 
(counterexample) state $\vx_{j+1} \in S$ for which the CLF condition fails. Assume without loss
of generality that $\vx_{j+1} \not= \vzero$.
\item Failing verification, the demonstrator is called to choose a suitable control $\vu_{j+1}$ corresponding to $\vx_{j+1}$.
\item The new set $W_{j+1} := W_j \cup \{ (\vx_{j+1}, \vu_{j+1}) \}$. Furthermore, 
\begin{equation}\label{eq:C_j+1}
	C_{j+1}:\ C_j \cap \left\{ \vc \ |\ \begin{array}{c} V_{\vc}(\vx_{j+1}) > 0\ \land\ \\\nabla V_{\vc}\cdot f(\vx_{j+1}, \vu_{j+1}) < 0 \end{array} \right\} \,.
\end{equation}
\end{compactenum}

We now prove some core properties that guarantee the correctness of
the proposed scheme. We assume that the learner and the verifier are
implemented without any approximations/relaxations (as will be
subsequently presented).

\begin{theorem}\label{thm:algo-learning-thm}
  The algorithmic learning framework as described above has the
  following property:
\begin{enumerate}
\item $\vc_j \not\in C_{j+1}$. I.e, the candidate found at the $j^{th}$ step is eliminated from further consideration.
\item If the algorithm succeeds at iteration $j$, then the output function $V_j(\vx)$ is a valid CLF for stabilization.
\item The algorithm declares failure at iteration $j$ if and only if no linear combination of the basis functions is a CLF compatible with the demonstrator. 
\end{enumerate}
\end{theorem}
\ifarxiv
\begin{proof}
	1) Suppose that $\vc_j \in C_{j+1}$. Then, $\vc_j$ satisfies the following conditions (Eq.~\eqref{eq:C_j+1}):
	\[
	V_{j}(\vx_{j+1}) > 0\ \land\ \nabla V_{j}\cdot f(\vx_{j+1}, \vu_{j+1}) < 0 \,.
	\]
	However, the verifier guarantees that $\vc_j$ is a counterexample for Eq.~\eqref{eq:clf-def}). I.e.
	\[
		V_{j}(\vx_{j+1}) \leq 0\ \lor\ \nabla V_{j}\cdot f(\vx_{j+1}, \vu_{j+1}) \geq 0 \,,
	\]
	which is a contradiction. Therefore, $\vc_j \not\in C_{j+1}$.
	
	2) The algorithm declares success if the verifier could not find a counterexample. In other words, $V_j$ satisfies conditions of Eq.~\eqref{eq:clf-def} and therefore a CLF.
	
	3) A CLF $V$ is compatible with a Demonstrator $\D$ iff
	\begin{equation}\label{eq:compatible-clf}
	(\forall \vx \in S \setminus \{\vzero\}) \ \nabla V\cdot f(\vx, \D(\vx)) < 0 \,.
	\end{equation}
	The algorithm declares failure if $C_j = \emptyset$ and Eq.~\eqref{eq:C_j} is equivalent to $C_j: C^1_{j} \cap C^2_{j}$ where
	\begin{align*}
	C^1_j :& \{ \vc \in C_0 | \land_{(\vx_i, \vu_i)\in W_j} \nabla V_\vc\cdot f(\vx_i, \vu_i) < 0 \} \\
	C^2_j :& \{ \vc \in C_0 | \land_{(\vx_i, \vu_i)\in W_j} V_\vc(\vx_i) > 0 \}
	\end{align*}
	 Suppose there exists a $\vc \in C_0$ (a linear combination of the basis function) that yields a CLF $V_\vc$. If $V_\vc$ is compatible with $\D$, then
	\[
	(\forall (\vx_i, \vu_i = \D(\vx_i)) \in W_j) \ \nabla V_{\vc}\cdot f(\vx_i, \vu_i) < 0 \,.
	\]
	Therefore, $\vc \in C^1_j$. But we know that $\vc \not\in C_j = \emptyset$. As a result $\vc \not\in C^2_j$. I.e.
	\[
	(\exists (\vx_i, \vu_i) \in W_j) \ V_{\vc}(\vx_i) \leq 0,
	\]
	and it is concluded that $V_\vc$ is not a CLF, which is a contradiction.
\end{proof}
\else
Proofs are available in~\cite{ravanbakhsh2017learning}. 

\fi
	Inverse results~\cite{peet2008polynomial} suggest polynomial basis for Lyapunov functions are expressive enough for verification of exponentially stable, smooth nonlinear systems. This, justifies using polynomial basis for CLF.
\begin{lemma}\label{lem:completeness}
	Assuming (i) the demonstrator function $\D$ is smooth, (ii) the closed loop system with controller $\D$ is exponentially stable, then there exists a polynomial CLF, compatible with $\D$.
\end{lemma}
\ifarxiv
\begin{proof}
	Under assumption (i) and (ii), one can show that a polynomial Lyapunov function $V$ (not control Lyapunov function) exists for the closed loop system $\Psi(X, U, \Plant , \D)$~\cite{peet2008polynomial}. I.e.
	\begin{align*}
		& V(\vzero) = 0 \\
		(\forall \vx \in S \setminus \{\vzero\}) \ & V(\vx) > 0 \\
		(\forall \vx \in S \setminus \{\vzero\}) \ & \nabla V\cdot f(\vx, \D(\vx)) < 0 \\
	\end{align*}
	$V$ is also a CLF since it satisfies Eq.~\eqref{eq:clf-def}. Moreover, $V$ is compatible with $\D$ (Eq.~\eqref{eq:compatible-clf}).
\end{proof}
\fi
Lemma~\ref{lem:completeness} guarantees success of the learning procedure if the set of basis functions is rich enough.
We now present implementations of each of the modules involved, starting with the learner.
\subsection{Learner: Finding a Candidate}
The $\textsc{findCandidate}$ function simply samples a point from the set 
$C_j$ defined in Eq.~\eqref{eq:C_j}.
Note that $V_{\vc}(\vx_j) : \vc^t \cdot \vg(\vx_j)$ is linear in $\vc$ and therefore $\nabla V_\vc.f(\vx_j, \vu_j)$ is linear in $\vc$ as well. The initial space of all candidates $C$ is assumed to be a hyper-rectangular open box. 
At each iteration, the candidate $\vc_j \in C_j$ is chosen.  Suppose
the algorithm does not terminate at this iteration. 
According to Eq.~\eqref{eq:C_j+1}, the new set $C_{j+1}$ is
obtained as $C_{j+1}:\ C_j \cap H_j$,
wherein $H_j$ is defined by two linear inequalities $H_{j1} \land H_{j2}$
($H_{j1}: \va_{j1}^t \vc < b_{j1}$, $H_{j2}: \va_{j2}^t \vc < b_{j2}$).
Let $\overline{C_j}$ represent the topological closure of the set
  $C_j$.
\begin{lemma}\label{lemma:cj-convex}
For each $j \geq 0$, $\overline{C_j}$ is a convex polyhedron.
\end{lemma}

\ifarxiv
\begin{proof}
	We prove by induction. Initially $C_0$ is an open box and as a result $\overline{C_0}$ is a convex polyhedron. Now, assume $\overline{C_j}$ is a polyhedron. Recall that $C_{j+1}$ is defined as
	\[
	C_{j+1}:\ C_j \cap \left\{ \vc \ |\ \begin{array}{c} \sum_{i=1}^{r} (c_i \ g_i(\vx_{j+1})) > 0\ \land\ \\ \sum_{i=1}^{r} (c_i \ \nabla g_i \cdot f(\vx_{j+1}, \vu_{j+1})) < 0 \end{array} \right\} \,.
	\]
	Notice that $f$ and $g_i$ are fixed $(\forall i)$ and $\vx_{j+1}$ and $\vu_{j+1}$ are given. Therefore, $C_{j+1}$ is intersection of an open polyhedron ($C_j$) and two half-space which yields an other polyhedron. And since the inequalities in equation above are strict, $C_{j+1}$ is an open polyhedron.
\end{proof}
\fi
Thus the problem of implementing \textsf{findCandidate} is that of
checking emptiness of a polyhedron with some strict inequality
constraints. This is readily solved using a slight modification of
standard linear programming algorithms using infinitesimals for strict
inequalities. 

\begin{figure}[t]
\begin{center}
\begin{tikzpicture}[scale=0.8]
\draw[draw=black, line width = 1.5pt, fill=green!20](0,0) -- (2,0.5) -- (2,2.5) -- (0,4) -- (-1,1) -- cycle;
\draw[fill=red!20] (0.5,1.45) circle (0.1);
\draw[draw=blue!30, line width=1.5pt, pattern = north west lines, pattern color=black] (0,0) -- (2,0.5) -- (2,1.1) -- (1.2,1.9) -- ( -0.35,0.35) -- cycle;
\draw[draw=black, line width=1.5pt, dashed](-0.7,0.0) -- (2.3, 3.0);
\draw[draw=black, line width=1.5pt, dashed](2.3,0.8) -- (-1, 4.1);
\node at (0.22,1.45) {$\vc_j$};
\node at (0.3,2.3) {$C_j$};
\node at (0.8,0.6) {$C_{j+1}$};
\node at (-1.,3.5){$H_{j1}$};
\node at (2.8,2.8){$H_{j2}$};
\end{tikzpicture}
\end{center}
\caption{Original candidate region $C_j$ (green) at the start of the
  $j^{th}$ iteration, the candidate $\vc_j$, and the new region
  $C_{j+1}$ (hatched region with blue lines). 
  Also, $\va_{j2} \vc_j > b_{j2}$ ($H_{j2}$ passes through $\vc_j$)}\label{fig:learning-iteration}
\vspace*{-0.5cm}
\end{figure}
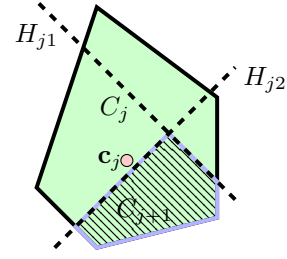

\begin{lemma}\label{lemma:cj-half-space}
There exists a halfspace  $H_{jk}:\ \va_{jk}^t \vc_j \geq b_{jk}$ that passes through $\vc_j$, and $C_{j+1} \subseteq C_j \cap H_{jk}$.
\end{lemma}

\ifarxiv
By definition $\vc_j \in C_j$. By Theorem~\ref{thm:algo-learning-thm}, it is guaranteed that $\vc_j \not\in C_{j+1}$. Since $C_{j+1} : C_j \cap \{\vc | a_{j1}^t \vc_j < b_{j1} \land a_{j2}^t \vc_j < b_{j2}\}$, we can conclude $\vc_j \not\in \{\vc | a_{j1}^t \vc_j < b_{j1} \land a_{j2}^t \vc_j < b_{j2}\}$, I.e.$a_{j1}^t \vc_j \geq b_{j1} \lor a_{j2}^t \vc_j \geq b_{j2}$. Thus one of the two disjuncts holds for $ k \in {1,2}$.
Further, we note that  $C_{j+1} \subseteq C_{j} \cap \{\vc | a_{jk}^t \vc_j < b_{jk} \}$.
\fi

\paragraph{Choosing a Candidate $\vc_j$:} The choice of $\vc_j \in C_j$ governs
the overall convergence of the algorithm. Figure~\ref{fig:learning-iteration}
demonstrates the importance of this choice by showing the candidate $\vc_j$,
the hyperplane $H_{jk}$ and the new region $C_{j+1}$.

We wish to choose $\vc_j$ s.t. $\mathsc{Vol}\left(C_{j+1}\right) \leq \alpha \mathsc{Vol}\left(C_{j}\right)$,
for a fixed $\alpha$, independent of $H_{jk}$.

\begin{theorem}[Tarasov et al.\cite{tarasov1988method}] \label{thm:mve}
Let $\vc_j$ be chosen as the center of the maximum volume ellipsoid (MVE) of $C_j$. Then,
\[ \mathsc{Vol}\left(C_{j+1}\right) \leq \left(1-\frac{1}{r}\right) \mathsc{Vol}\left(C_{j}\right) \,.\]
\end{theorem}

This leads us to a  scheme that guarantees termination
of the overall iterative scheme in finitely many steps under a
robustness assumption.

\begin{definition}[Robust Candidate]
A candidate $\vc \in C$ is $\delta$-robust ($\delta > 0$), iff 
for each $\hat{\vc} \in \B_{\delta}(\vc)$, $V_{\hat{\vc}}:\hat{\vc}^t \cdot \vg(\vx)$ is
a CLF, where $\B_\delta(\vc)$ is a ball of 
radius $\delta$ around $\vc$.
\end{definition}

Let $\mathsc{Vol}(C) < \gamma \Delta^r$ for $\gamma > 0$ the volume of
the unit $r$-ball, and $\Delta > 0$. Furthermore, the
procedure terminates
whenever $\mathsc{Vol}(C_j) < \gamma \delta^r$ following the robustness
assumption above.
\begin{theorem} \label{thm:termination}
If at each step $\vc_j$ is chosen as the center of the maximum volume ellipsoid in $C_j$, the learning loop terminates in at most
\[ \frac{r (\log(\Delta) - \log(\delta))}{- \log\left(1 - \frac{1}{r}\right) } = O(r^2) \ \mbox{iterations}\,.\]
\end{theorem}

\ifarxiv
\begin{proof}
	Initially, $\mathsc{Vol}(C_0) < \gamma \Delta^r$. Then by Theorem~\ref{thm:mve}
	\begin{align*}
	&\mathsc{Vol}(C_j) \leq (1 - \frac{1}{r})^j \ \mathsc{Vol}(C_0) < (1 - \frac{1}{r})^j \gamma \Delta^r \\
	\implies & \log(\mathsc{Vol}(C_j)) - \log(\gamma \Delta^r) < j \  \log(1-\frac{1}{r})
	\end{align*}
	Suppose the algorithm does not terminate after $\frac{r(\log(\Delta)-\log(\delta))}{-\log(1-\frac{1}{r})}$ iterations. Then
	\begin{equation*}
		\log(\mathsc{Vol}(C_j)) - \log(\gamma \Delta^r) < \frac{r(\log(\Delta)-\log(\delta))}{-\log(1-\frac{1}{r})} \  \log(1-\frac{1}{r})
	\end{equation*}
	and
	\begin{align*}
		\implies & \log(\mathsc{Vol}(C_j)) - \log(\gamma \Delta^r) < r(\log(\delta) - \log(\Delta)) \\
		\implies & \log(\mathsc{Vol}(C_j)) - \log(\gamma \Delta^r) < \log(\gamma \delta^r) - \log(\gamma \Delta^r) \\
		\implies & \log(\mathsc{Vol}(C_j)) < \log(\gamma \delta^r)
	\end{align*}	
	And it is concluded that $\mathsc{Vol}(C_j) < \gamma \delta^r$, which is the termination condition. This is a contradiction and therefore, the algorithm terminates in at most $\frac{r(\log(\Delta)-\log(\delta))}{-\log(1-\frac{1}{r})}$
	iterations. And asymptotically $-\log(1 - \frac{1}{r})$ is $\Omega(\frac{1}{r})$ (can be shown using Taylor expansion as $r \rightarrow \infty$) and therefore, the maximum number of iterations would be $O(r^2)$.
\end{proof}
\fi

The maximum volume ellipsoid itself can be computed by solving a convex optimization problem\cite{vandenberghe1998determinant}.

\subsection{Implementing the Verifier}

The verifier given a candidate $V_j(\vx): \vc_j^t \cdot \vg(\vx)$ checks the CLF conditions in Eq.~\eqref{eq:clf-def}, split into two separate
checks:

\noindent\textbf{(A)} Check if $V_j(\vx)$ is a positive definite polynomial over
  $\vx \in S$. Assuming that $\vg(\vzero) =0$ for the basis, this
  reduces to:
  \begin{equation}\label{eq:positivity-cond}
 (\exists\ \vx \in S \setminus \{\vzero\})\ V_j(\vx) \leq 0 \,.
\end{equation}

\noindent\textbf{(B)} Check if the Lie derivative of $V_j$ can be made negative for
each $\vx \in S$ by a choice $\vu \in U$: 
\begin{equation}\label{eq:decrease-cond-init}
 (\exists \vx \in S  \setminus \{\vzero\} ) \ (\forall \vu \in U)\ (\nabla V_j) \cdot f(\vx, \vu) \geq 0 \,.
\end{equation}
This problem \emph{seems} harder  due to the presence of a \emph{quantifier alternation}. Let $U$ be a polyhedral set defined by $U:\ \{ \vu \in \reals^{m}\ |\  A \vu \geq \vb \}$.
Recall that $f(\vx, \vu)$ is control affine function $f_0(\vx) + \sum_{i=1}^m f_i(\vx) u_i$.
\begin{lemma}
Eq.~\eqref{eq:decrease-cond-init} holds for some $\vx \in S$ iff
\begin{equation}\label{eq:decr-condition}
\begin{array}{ll}
(\exists\ \vx\in S\setminus\{\vzero\},\vlam) & \hspace{-0.2cm}  \vlam\geq \vzero, \vlam^t \vb \geq -\nabla V_j.f_0(\vx) \\	
& \hspace{-0.2cm} A_i^t \vlam=\nabla V_j.f_i(\vx) (i \in \{1 \ldots m\}).
\end{array}
\end{equation}
\end{lemma}

\ifarxiv
\begin{proof}
	Considering a fixed $V$ and $\vx$, then
	\begin{equation} \label{eq:u-cond-before-farkas-lemma}
	(\forall \vu \in U) \ \nabla V \cdot f(\vx, \vu) = \nabla V \cdot f_0(\vx) + \sum_{i=1}^{m} \nabla V \cdot f_i(\vx) u_i \geq 0 \,,
	\end{equation}
	which is equivalent to:
	\[
		(\not\exists \vu) A \vu \geq \vb \land \nabla V \cdot f_0(\vx) + \sum_{i=1}^{m} \nabla V \cdot f_i(\vx) u_i < 0 \,.
	\]
	 This yields a set of linear inequalities (w.r.t. $\vu$). Using Farkas lemma, it is concluded
	\[
	(\exists \vlam \geq 0) \ A_i^t \vlam = \nabla V \cdot f_i(\vx) (i \in \{1...m\}), \vlam^t \vb \geq -\nabla V \cdot f_0(\vx).
	\]
\end{proof}
\fi

In other words, assuming that the dynamics and chosen bases are
polynomials, the verification problem reduces to checking if a given
semi-algebraic set defined by polynomial inequalities has a solution.

Failing the polynomial assumption, the problem of verification is
\emph{in general} undecidable. However, it can be approximated by
techniques such as $\delta$-decision procedures proposed by Gao et
al~\cite{gao2013dreal}. Solvers like dReal can thus be directly used
for the verification problem. While dReal does a good job in adaptive space decomposition, in our experience, they do not scale \emph{reliably}. Nevertheless,  these solvers
allow us to conveniently implement a verifier for small but hard
problems involving rational and trigonometric functions.

The verification problem for polynomial systems and polynomial CLFs
through a polynomial basis function $\vg(\vx)$ is NP-hard, in
general. Exact approaches using semi-algebraic geometry and the
branch-and-bound solvers (including the dReal approach cited above)
can tackle this problem precisely. However, scalability is still an issue.

We sketch a relaxation using SDP solvers~\cite{mosek2010mosek}.
Let us fix a basis of monomial terms of degree up to $D_V$,
$\vm^t: [1 \  x_1 \ \ldots \ \vx^{D_V}]$, wherein $D_V$ is chosen as at least half of the maximum
degree in $\vx$ among all monomials in $g_j(\vx)$ and
$\nabla g_j \cdot f(\vx)$:
$D_V \geq \frac{1}{2} \max\left( \bigcup_{j} \left( \{ \mbox{deg}(g_j) \} \cup \{ \mbox{deg}(\nabla g_j \cdot f ) \} \right) \right).$
Let $Z(\vx): \vm \vm^t$. Thus, each polynomial of degree up to $2D_V$ may now be written as
a trace inner product $p(\vx) =\ \tupleof{ P, Z(\vx)} = \mathsf{trace}( P Z(\vx) )$, wherein the matrix $P$ is constant.

Let $S$ be the semi-algebraic set defined as 
\[ S: \{ \vx \in \reals^n\ |\ r_1(\vx) \leq 0, \ldots, r_k(\vx) \leq 0 \} \,,\]
for polynomials $r_1, \ldots, r_k$.
The constraint in ~\eqref{eq:positivity-cond}
is equivalent to solving the following optimization
problem over $\vx$
\begin{equation}\label{eq:positivity-cond-relax}
\begin{array}{ll}
 \mathsf{max}_{\vx} \tupleof{I,Z(\vx)} & \\
 \mathsf{ s.t. }  & \tupleof{R_i, Z(\vx)} \leq 0,\ i \in \{ 1,\ldots, k \} \\
  & \tupleof{\mathcal{V}_j, Z(\vx)} \leq 0 \,, \\
\end{array}
\end{equation}
where $V_j(\vx)$ ($r_i(\vx)$) is written as
$\tupleof{\mathcal{V}_j, Z(\vx)}$ ($\tupleof{R_i, Z(\vx)}$).
and those in ~\eqref{eq:decr-condition} are written as
\begin{equation}\label{eq:decr-cond-relax}
 \begin{array}{ll}
 \mathsf{max}_{\vx, \vlam} \tupleof{I,Z(\vx)} & \\
 \mathsf{s.t.} & \hspace{-0.5cm} \tupleof{R_i, Z(\vx)} \leq 0,\ i \in \{ 1,\ldots, k \} \\
& \hspace{-0.5cm} \tupleof{F_{ji}, Z(\vx)} = A_i^t\vlam,\ i \in \{1,\ldots, m\}  \\
& \hspace{-0.5cm} \tupleof{-F_{j0}, Z(\vx)} \leq \vb^t \vlam , \ \vlam \geq 0 \,, \\
\end{array}
\end{equation}
wherein the components $\nabla V_j \cdot f_i(\vx)$ 
defining the Lie derivatives of $V_j$ are now written
in terms of $Z(\vx)$ as $\tupleof{F_{ji},Z(\vx)}$.

The SDP relaxation is used to solve these problems and provide an upper
bound of the solution~\cite{henrion2009gloptipoly}. The result of the
relaxation treats $Z(\vx)$ as a matrix variable $Z$ that will satisfy
$Z \succeq 0$.  Notice that each optimization problem is feasible
simply by setting $Z$ and $\vlam$ to be zero. However, if the optimal
solution of both problems is $Z = 0$ in the SDP relaxation, then we
will conclude that the given candidate is a CLF.

\begin{lemma}
If the relaxed optimization problems in  Eqs.~\eqref{eq:positivity-cond-relax} and ~\eqref{eq:decr-cond-relax}
yield a zero solution,  then the given candidate $V_j(\vx)$ is in fact a CLF. 
\end{lemma}

\ifarxiv
\begin{proof}
	Suppose that $V_j$ is not a CLF. Then it does satisfies either Eq.~\eqref{eq:positivity-cond} or Eq.~\eqref{eq:decrease-cond-init}. 
	I.e. $(\exists \vx^* \neq \vzero)$ s.t. $V_{j}(\vx^*) \leq 0$ or $(\exists\ \vlam \geq \vzero) \ A_i^t \vlam=\nabla V_{j}.f_i(\vx^*) (i \in \{1 \ldots m\}) ,\vlam^t \vb \geq - \nabla V_j.f_0(\vx^*)$. 
Then $Z(\vx^*)$ is a non-zero solution for Eq.~\eqref{eq:positivity-cond-relax} or Eq.~\eqref{eq:decr-cond-relax}. $Z(\vx^*)$ can also yield a non-zero relaxed matrix $Z^*$ which satisfies the relaxed optimization problem.

On the other hand, For any non-zero $Z \succeq 0$, $\tupleof{I,Z}$ is positive. Therefore, since the optimization returns a zero as the solution, it means the relaxed optimization does not have a non-zero solution (since $\tupleof{I,Z}$ is being maximized). Which is a contradiction as it is shown non-zero solution exists when $V_j$ is not a CLF. Therefore, $V_j$ is a CLF.
\end{proof}
\fi

However, the converse is not true. It is possible for $Z \succ 0$ to
be optimal for either relaxed condition, but no 
$\vx \in \reals^n$ corresponds to the solution.  
This happens
because the relaxation drops two key constraints to convexify the
conditions: (1) $Z$ has to be a rank one matrix written as
$Z: \vm \vm^t$ and (2) there is a $\vx \in \reals^n$ such that $\vm$
is the matrix of monomials corresponding to $\vx$. 

To deal with this, we adapt our learning framework to 
work with witnesses $W_j: \{ (Z_i, \vu_i) \}_{i=1}^j$ replacing states
$\vx_i$ by matrices $Z_i$.

\begin{enumerate}
\item Each basis function $g_j(\vx)$ in $\vg$ is now written instead as $\tupleof{G_j, Z}$.
The candidates are therefore, $\sum_{j=1}^r  c_j \tupleof{ G_j, Z}$.
 Likewise, we write the components of its Lie
 derivative $\nabla g_j \cdot f_i$ in terms of $Z$.

\item The learner maintains the set $W$ as $\{ (Z_j, \vu_j) \}$, wherein $Z_j$ is the feasible
solution returned by the SDP solver while solving Eqs.~\eqref{eq:decr-cond-relax} and ~\eqref{eq:positivity-cond-relax}.
In other words, the CLF conditions are, in fact, taken to be the relaxed conditions.

\item We use a suitable projection operator $\pi$ mapping each $Z$ to
  a state $\vx: \pi(Z)$, such that the demonstrator receives $\vx$. In
  practice, since the vector of monomials used to define $Z$ from
  $\vx$ includes the terms $1, x_1, \ldots, x_n$, the projection
  operator simply selects a few entries from $Z$ corresponding to the
  variables. Other more sophisticated projections are also possible.

\end{enumerate}

The relaxed framework thus lifts counterexamples to work over matrices
$Z_j$. However, the candidate space begins with $C$ and is refined
each step as before. I.e, the relaxed framework continues to satisfy
Lemmas~\ref{lemma:cj-convex},~\ref{lemma:cj-half-space}, Theorem~\ref{thm:termination} and
Theorem~\ref{thm:algo-learning-thm} with the definition of (control)
Lyapunov function changed to relaxed conditions.

\section{Experiments}\label{sec:expr}
In this section, we describe numerical results on some example benchmark systems.
The algorithmic framework is implemented using quadratic template forms
for the CLFs with the tool Globoptipoly used to 
implement the verifier~\cite{henrion2005solving}. The demonstrator
is implemented using a nonlinear MPC implemented using a gradient
descent algorithm. For each benchmark, we tuned the time horizon, discretization step 
and the cost function until the control objectives were satisfied by the MPC over
hundreds of simulations starting from randomly selected initial states. 

Most of the benchmarks consider a \emph{reach-while-stay} problem,
wherein the goal is to reach target set $T$, starting initial set $I$,
while remaining in the safe set $S$.  We also illustrate an example
involving a trajectory stabilization problem. All the computations are
performed on a Mac Book Pro with 2.9 GHz Intel Core i7 processor and
16GB of RAM. The reported CLFs are rounded to $2$ decimal points.  The
summary of results is provided in Table.~\ref{tab:result}.

 \begin{table}[t]
\caption{\small Results on the numerical examples. $n$: \# variables, $m$: \# control inputs, $\tau$: MPC time step, $\T$: MPC time horizon, $D_V$: SDP relaxation degree bound for the \textsc{Verifier}, \#Itr: \# of iterations, Time: total computation time (minutes). }\label{tab:result} 
\begin{center}
\begin{tabular}{ |l|c|c||c|c|c|c|c| } 
 \hline
 System Name & $n$ & $m$ & $\tau$ & $\T$ & $D_V$ & \# Itr & Time \\
 \hline
 Tora & 4 & 1 & 1 & 30 & 4 & 53 & 30 \\ 
 Bicycle & 4 & 2 & 0.4 & 8 & 3 & 51 & 9 \\ 
 Bicycle $\times$ 2 & 8 & 4 & 0.4 & 8 & 3 & 536 & 303 \\ 
 Inverted Pendulum & 4 & 1 & 0.04 & 2 & 5 & 85 & 31 \\
 Forward Flight & 4 & 2 & 0.4 & 16 & 5 & 32 & 26 \\ 
 Hover Mode & 6 & 2 & 0.4 & 16 & 4 & 213 & 163  \\
 Unicycle (Seg.1) & 3+1* & 2 & 0.1 & 2 & 4 & 41 & 15\\
 Unicycle (Seg.2) & 3+1* & 2 & 0.1 & 3 & 4 & 31 & 7 \\
 \hline
\end{tabular}
\\ * +1 refers to the time variable.
\end{center}
\end{table}

 \begin{figure}[t]
\begin{center}
	\includegraphics[width=0.4\textwidth]{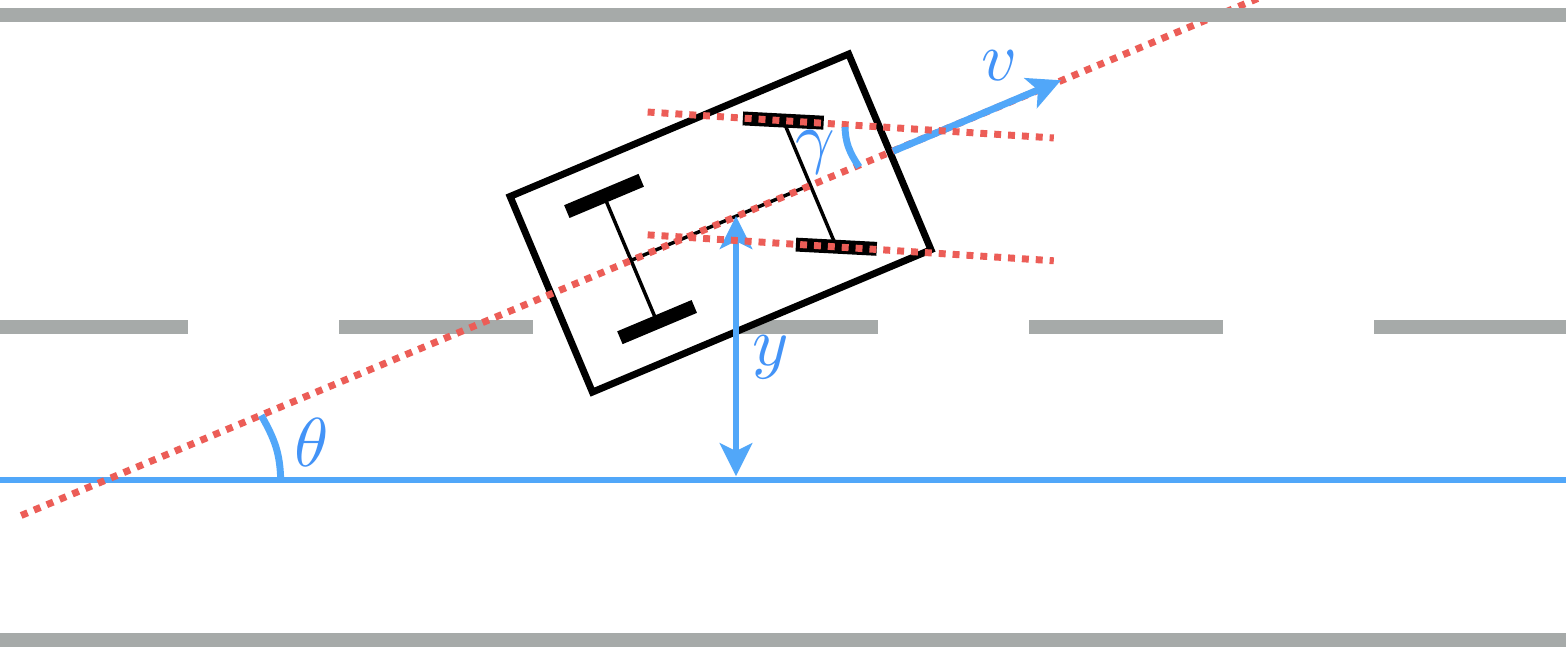}
\end{center}
\caption{A Schematic View of the Bicycle Model for Stabilizing to the Road.}\label{fig:bicycle} 
\end{figure}
\paragraph{Bicycle Model:} This system is two-wheeled mobile robot
modeled with five states $[x, y, v, \theta, \gamma]$ and two control
inputs~\cite{francis2016models}.  The goal is to stabilize the car 
to a target velocity $v^*=5$, as
shown in Fig.~\ref{fig:bicycle}. We drop the variable $x$ (since it is
immaterial to our stabilization problem) and obtain a model with four state
variables:
\begin{equation*}\label{ex:bicycle-dyn}
	\left[ \begin{array}{l}
		\dot{y} \\ \dot{v} \\ \dot{\theta} \\ \dot{\sigma}
	\end{array}\right] = 
	\left[ \begin{array}{l}
		v\sin(\theta) \\ u_1 \\ v\sigma \\ u_2
	\end{array} \right] \,,
	\begin{array}{ll}
		U: [-10, 10]\times[-10, 10] \\
		S: [-2, 2]\times[3, 7]\times[-1, 1]^2 \\
		I: \B_{0.4}(\vzero) \\
		T: \B_{0.1}(\vzero) \,,
	\end{array}
\end{equation*}
where $\sigma = tan(\gamma)$ (see Fig.~\ref{fig:bicycle}). 
$\sin$ function is approximated with a
polynomial of degree $1$. The method finds the following CLF:
\begin{align*}
V =
&+ 0.42 y^2
+ 0.59 y\theta
+ 2.57 \theta^2
+ 0.79 y\sigma
+ 4.64 \sigma\theta\\
&+ 4.06 \sigma^2
- 0.38 vy
+ 1.46 v\theta
+ 1.18 v\sigma
+ 2.39 v^2\,.
\end{align*}

\paragraph{Inverted Pendulum on a Cart:} This example has applications in
balancing two-wheeled robots~\cite{chan2013review} (cf.~\cite{anderson2002nbot}
for list of such robots). The system has four
state variables $[x, \dot{x}, \theta, \dot{\theta}]$ and one input $u$.
The dynamics, after partial
linearization, have the following form~\cite{landry2005dynamics}:

\begin{equation*}\label{eq:inverted-pendulum-dyn}
	\left[ \begin{array}{l}
		\ddot{\vx} \\ \ddot{\theta}
	\end{array}\right] = 
	\left[ \begin{array}{l}
		4u + \frac{4(M+m)g \tan(\theta) - 3mg\sin(\theta)\cos(\theta)}{4(M+m)-3m\cos^2(\theta)} \\ \frac{- 3 u \cos(\theta)}{l}
	\end{array} \right] \,,
\end{equation*}
where $m = 0.21$, $M=0.815$, $g=9.8$ and $l=0.305$.
The trigonometric and rational functions are approximated with polynomials of degree $3$. The sets are $ S: [-1, 1]^4, U:[-20, 20], I:\B_{0.1}(\vx), T: \B_{0.1}(\vx)$. We obtain the following CLF:
\begin{align*}
V =& + 11.64 \dot{\theta}^2 + 45.93 \dot{\theta}\theta
+ 85.47 \theta^2 + 12.15 x\dot{\theta} 
+ 36.57 x\theta \\
&+ 6.44 x^2 + 15.07 \dot{\theta}\dot{x} + 33.06 \dot{x}\theta
+ 8.98 \dot{x}x + 6.09 x^2 \,.
\end{align*}

\paragraph{Forward Flight for Caltech Ducted Fan:} The Caltech ducted
fan models the aerodynamics of a single wing of a thrust vectored,
fixed wing aircraft~\cite{jadbabaie2002control}.  This problem is to
design forward flight control in which the angle of attack needs to be
changed. The model of the system is carefully calibrated through wind tunnel experiments. The system has $4$ states
$[v, \gamma, \theta, q]$ and two control inputs: $u$ and $\delta_u$.
The dynamics are:
\begin{equation*}\label{ex:ducted-fan-forward-dyn}
	\left[ \begin{array}{l}
		m \dot{v} \\ m v \dot{\gamma} \\ \dot{\theta} \\ J \dot{q}
	\end{array}\right] = 
	\left[ \begin{array}{l}
		-D(v, \alpha) - W \sin(\gamma) + u \cos(\alpha + \delta_u) \\
		L(v, \alpha) - W \cos(\gamma) + u \sin(\alpha + \delta_u) \\
		q \\ M(v, \alpha) - u l_T \sin(\delta_u)
	\end{array} \right] \,,
\end{equation*}
where $\alpha = \theta - \gamma$, and $D$, $L$, and $M$ are polynomials in $v$ 
and $\alpha$. For full list of parameters, see~\cite{jadbabaie2002control}. The goal is to reach $\vx^*:\ [6, 0, 0.1771, 0]$ as the steady state.
We perform a translation so that
the $\vx^*$ is the origin in the new coordinate system.
\begin{align*}
U &:[0, 13.5] \times [-0.45, 0.45]\\
S &:[3, 9]\times[-0.75, 0.75]\times[-0.75, 0.75]\times[-2, 2]] \\
I &: \{[v, \gamma, \theta, q] |(4v)^2 + (10\gamma)^2 + (10\theta)^2 + (10q)^2 < 4^2 \} \, \\
T &: \{[v, \gamma, \theta, q] |(4v)^2 + (10\gamma)^2 + (10\theta)^2 + (10q)^2 < 0.5^2 \} \,.
\end{align*}

First, we approximate $v^{-1}$, $\sin$ and $\cos$ with polynomials of degree $1$, $3$ and $3$ respectively, to get polynomial dynamics. However, the system is not affine in control. We replace $u$ and $\delta_u$ input variables with $u_1 = u \sin(\delta_u)$
and $u_2 = u \cos(\delta_u)$ and $U$ is under-approximated with a polytope.
These changes yield a polynomial control affine dynamics, which fits the
description of our model. The procedure finds
the following CLF:
\begin{align*}
V =&+ 3.21 q^2 + 2.18 q\theta
+ 3.90 \theta^2 + 0.40 qv
- 0.15 v\theta \\
& + 0.56 v^2 + 1.78 q\gamma - 1.42 \gamma\theta
- 0.11 v\gamma + 3.90 \gamma^2 \,.
\end{align*}

\paragraph{Hover Mode for Caltech Ducted Fan:} This problem is again taken from~\cite{jadbabaie2002control}.
The goal is to keep the ducted fan in a hover mode. The system 
has $6$ variables $x$, $y$, $\theta$, $\dot{x}$, $\dot{y}$, $\dot{\theta}$ and two control inputs $u_1$, $u_2$. The dynamics are
\begin{equation*}\label{ex:ducted-fan-hover-dyn}
	\left[ \begin{array}{l}
		m \ddot{x} \\ m \ddot{y} \\ J \ddot{\theta}
	\end{array}\right] = 
	\left[ \begin{array}{l}
		-d_c\dot{x} + u_1 \cos(\theta) - u_2 \sin(\theta) \\
		-d_c\dot{y} + u_2 \cos(\theta) + u_1 \sin(\theta) - mg \\
		r u_1
	\end{array} \right] \,,
\end{equation*}
where $m = 11.2$, $g = 0.28$, $J = 0.0462$, $r = 0.156$ and $d_c = 0.1$. The sets are:
\begin{align*}
S &: [-1,1]\times[-1,1]\times[-0.7,0.7]\times[-1, 1]^3 \\
U &: [-10, 10]\times[0, 10], I:\B_{0.25}(\vzero), T:\B_{0.05}(\vzero) \,.
\end{align*}

The $sin$ and $cos$ are approximated with
polynomials of degree $2$ and the procedure finds a quadratic CLF:
\begin{align*}
V =& 1.63 \dot{\theta}^2 - 1.09 \dot{\theta}\dot{y}
+ 15.00 \dot{y}^2 - 0.02 \dot{\theta}y + 1.54 y\dot{y} 
+ 1.25 y^2 \\
&+ 1.74 \theta\dot{\theta} + 0.79 \dot{y}\theta + 0.08 y\theta + 4.86 \theta^2 - 4.93 \dot{\theta}\dot{x} \\
&+ 0.57 \dot{x}\dot{y} + 0.05 y\dot{x} - 8.26 \dot{x}\theta + 12.58 \dot{x}^2 - 0.44 \dot{\theta}x \\
&- 0.38 \dot{y}x - 0.20 yx 
- 4.27 x\theta + 3.86 x\dot{x} + 2.14 x^2 \,. & 
\end{align*}

\paragraph{Unicycle:} Now, we consider a simpler system, namely
the unicycle model~\cite{liberzon2012switching}, which is known not to have 
a polynomial CLF. However,
for trajectory tracking problem, one can provide a time varying 
CLF~\cite{tedrake2010lqr}
for a finite time horizon control (using funnels). 
The unicycle model has the dynamics: $\dot{x_1} = u_1, \dot{x_2} = u_2,  \dot{x_3} = x_1 u_2 - x_2 u_1$.

\begin{figure}[t]
\begin{center}
	\includegraphics[width=0.35\textwidth]{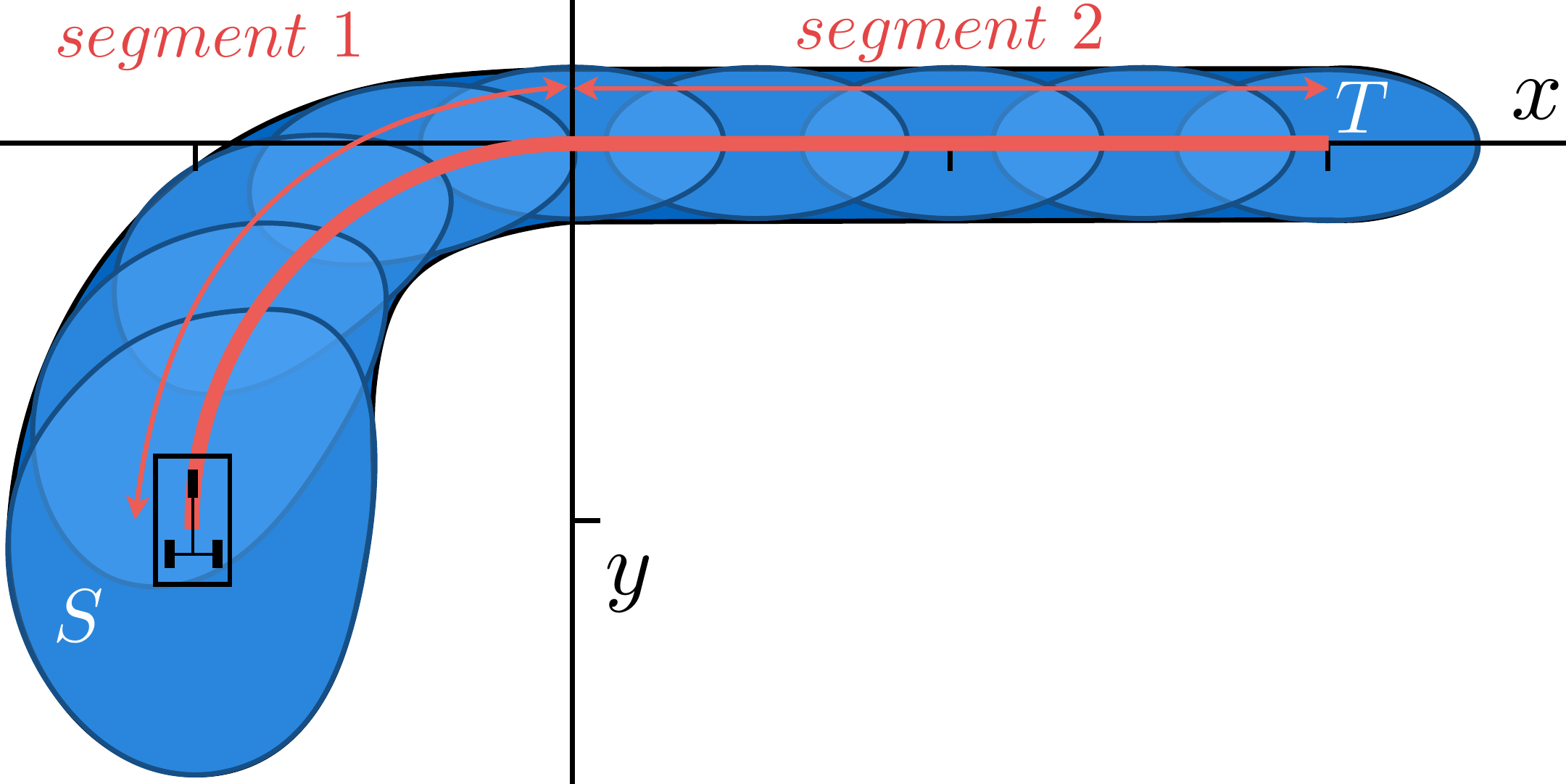}
\end{center}
\caption{Trajectory Tracking using Time-varying CLF - Projected on x-y
 plane. The reference trajectory is shown with the red line, consists of two segments.
 Starting from $S$, the state remains in the funnel (blue region) 
 until it reaches $T$. Boundary of each smaller blue region shows
 a level-set of $V$ for a specific time.
 }\label{fig:unicycle} 
\end{figure}

We consider a planning problem: starting
near $[\frac{\pi}{2}, -1, -1]$, the goal is to reach near $[0, 2, 0]$, and by
near we mean within distance $1$.
In the first step, a feasible trajectory $\vx(t)$ is generated as shown in Fig.~\ref{fig:unicycle}. 
Then $\vx(t)$ is approximated with piecewise polynomials. More precisely, trajectory consists
 of two segments. The first segment brings the car to the
origin and the second segment moves the car to the
destination. Each segment is approximated using
polynomials in $t$ with degree up to $3$. 
The time varying CLF $V$ is a function of both 
the state $\vx$ and time $t$.
Choosing a template for $V$, our method is applied to this problem
and we are able to find a strategy to implement the plan with guarantees.
A level-set of the funnel (CLF) is shown in Fig.~\ref{fig:unicycle}.

\section{Related Work}\label{section:related-work}

\paragraph{Synthesis of Lyapunov Functions from Data:} The problem of
synthesizing Lyapunov functions for a control system by observing the
states of the system in simulation has been investigated in the past
by Topcu et al to learn Lyapunov functions along with the resulting
basin of attraction~\cite{topcu2007stability}. Whereas the original
problem is bilinear, the use of simulation data makes it easier to
postulate states that belong to the region of attraction and therefore
find Lyapunov functions that belong to this region by solving LMIs in
each case. The application of this idea to larger black-box systems is
demonstrated by Kapinski et al~\cite{kapinski2014simulation}. Our
approach focuses on controller synthesis through learning a control
Lyapunov function in a bid to replace an existing controller. In doing
so, we do not attempt to prove that the original demonstrator is
necessarily correct but find a control Lyapunov function by assuming
that the demonstrator is able to stabilize the system for the initial
conditions we query on. Another important contribution lies in our
analysis of the convergence of the learning with a bound on the
maximum number of queries needed. In fact, these results can also be
applied to the Lyapunov function synthesis approaches mentioned
earlier.

\paragraph{Counter-Example Guided Inductive Synthesis (CEGIS):} Our
approach of alternating between a learning module that proposes a
candidate and a verification module that checks the proposed candidate
is identical to the CEGIS framework originally proposed by
Solar-Lezama et al.~\cite{solar2006combinatorial,solar2008program}.
As such, the CEGIS approach does not include a demonstrator that can
be queried. The extension of this approach Oracle-guided inductive
synthesis~\cite{jha2015theory}, generalizes CEGIS using an
\emph{oracle} that serves a similar role as a demonstrator in this
paper. Jha et al.  prove bounds on the number of queries for discrete
concept classes using results on exact concept learning in discrete
spaces~\cite{goldman1995complexity}.  

The CEGIS procedure has been used for the synthesis of CLFs recently
by authors~\cite{ravanbakhsh2015counter,ravanbakhsh2016robust}, combining it
with SDP solvers for verifying CLFs and a robust version for switched
systems. The key difference here lies in the use of the demonstrator
module that simplifies the learning module. In the absence of a
demonstrator module, the problem of finding a candidate reduces to
solving linear constraints with disjunctions, an NP-hard
problem~\cite{ravanbakhsh2016robust}. Likewise, the convergence
results are quite weak~\cite{ravanbakhsh2015counter}. In the setting
of this paper, however, the use of a MPC scheme as a demonstrator
allows us to use faster LP solvers and provide convergence guarantees.

\paragraph{Learning from Demonstration:} The idea of learning from
demonstration has had a long history in
robotics~\cite{argall2009survey}. The overall framework uses a
demonstrator that can in fact be a human operator~\cite{khansari2017learning} or a complex
MPC-based control law~\cite{kahn2016plato,zhang2015learning,zhong2013value,mordatch2014combining}. 
The approaches differ on the nature of the
interaction between the learner and the demonstrator; as well as how
the policy is inferred. Our approach stands out in many ways: (a) We
represent our policies by CLFs which are polynomial. On one hand,
these are much less powerful than approaches that use neural
networks~\cite{zhang2015learning,kahn2016plato}, for instance. However, the
advantage lies in our ability to solve verification problems to ensure
that the resulting policy learned through the CLF is correct with
respect to the underlying dynamical model. (b) Our framework is
\emph{adversarial}. The choice of the counterexample to query the
demonstrator comes from a failed attempt to validate the current
candidate. (c) Finally, we use simple yet powerful ideas from convex
optimization to place bounds on the number of queries, paralleling
some  results on concept learning in discrete
spaces~\cite{goldman1995complexity}.

Control Lyapunov functions were originally introduced by Artstein and
the construction of a feedback law given a CLF was first given by
Sontag~\cite{artstein1983stabilization,sontag1989universal}. As such,
the problem of learning CLFs is well known to be hard, involving
bilinear matrix inequalities (BMIs)~\cite{tan2004searching}. An
equivalent approach involves solving bilinear problems simultaneously
for a control law and a Lyapunov function certifying
it~\cite{el1994synthesis,majumdar2013control}. BMIs are well known to
be NP-hard, and hard to solve for a feasible
solution~\cite{henrion2005solving}. The common approach is to perform
an alternating minimization by fixing one set of bilinear variables
while minimizing in the other. Such an approach has poor guarantees in
practice, often ``getting stuck'' on a saddle point that does not
allow the technique to make progress in finding a feasible
solution. To combat this, Majumdar et al. (ibid) use LQR controllers
and their associated Lyapunov functions for the linearization of the
dynamics as good initial seed solutions~\cite{majumdar2013control}. In
contrast, our approach simply assumes a demonstrator in the form of a
MPC controller that can be used to resolve the
bilinearity. Furthermore, our approach does not encounter the local
saddle point problem.

The use of the learning framework with a demonstrator distinguishes
the approach in this paper from recently developed ideas based on
formal synthesis~\cite{wongpiromsarn2011tulip,liu2013synthesis,rungger2016scots,mouelhi2013cosyma,ozay2013computing,kloetzer2008fully,taly2011synthesizing,ravanbakhsh2015counter,ravanbakhsh2016robust,huang2015controller}. These
techniques focus on a given dynamical system and a specification of
the correctness in temporal logic to solve the problem of controller
design to ensure that the resulting trajectories of the closed loop
satisfy the temporal specifications. Majority of the approaches are
based on discretization of the state-space into cells to compute a
discrete abstraction of the overall
system~\cite{wongpiromsarn2011tulip,liu2013synthesis,rungger2016scots,mouelhi2013cosyma,ozay2013computing,kloetzer2008fully}. A
smaller set of approaches synthesize Lyapunov-like functions by
solving nonlinear constraints either through branch-and-bound
techniques or by sum-of-square (SOS) relaxation
techniques~\cite{taly2011synthesizing,ravanbakhsh2015counter,ravanbakhsh2016robust}.

In this paper, we use the Lyapunov function approach to synthesizing
controllers. An alternative is to use occupation
measures~\cite{rantzer2001dual,prajna2004nonlinear,lasserre2008nonlinear,majumdar2014convex}.
These methods formulate an infinite dimensional problem to maximize
the region of attraction and obtain a corresponding control law. This
is relaxed to a sequence of finite dimensional
SDPs~\cite{lasserre2001global}. Note however that the approach
computes an over approximation of the finite time backward reachable
set from the target and a corresponding control. Our framework here
instead seeks an under approximation that yields a guaranteed
controller.

\section{Conclusion And Future Work} 
\label{sec:conclusion}
We have proposed an algorithmic learning framework for synthesizing
CLFs using a demonstrator and demonstrated our approach on challenging
numerical examples with 4-8 state variables.  As future work, we are
considering many directions including the extensions to
noisy/erroneous demonstrations, using output feedback (rather than
full state feedback) synthesis and allowing disturbances in our
framework. We are also working on integrating our control framework
with RRT-based path planning and implementing it on board robotic
vehicles.

\section*{Acknowledgments}
This work was funded in part by NSF under award numbers SHF 1527075 and CPS 1646556. All opinions expressed
are those of the authors and not necessarily of the NSF.


\bibliographystyle{plainnat}
\bibliography{references}

\end{document}